\documentclass[prl,twocolumn,showpacs,twoside,showpacs]{revtex4}

\usepackage{graphicx}
\usepackage{dcolumn}
\usepackage{bm}
\usepackage{amsmath}

\begin{document}

\bibliographystyle{prsty}
\title{Neutron star structure and collective excitations of finite nuclei}
\author{N. Paar$^1$}
\email{npaar@phy.hr}
\author{Ch. C. Moustakidis $^{2}$}
\author{T. Marketin$^{1}$}
\author{D. Vretenar$^1$}
\author{G. A. Lalazissis $^{2}$}
\affiliation{$^1$Physics Department, Faculty of Science, University of Zagreb, 
Croatia}
\affiliation{$^2$Department of Theoretical Physics, Aristotle University of Thessaloniki, GR-54124 Thessaloniki, Greece}

\date{\today}
\begin{abstract}

We study relationships between properties of collective excitations in finite nuclei and the
phase transition density $n_t$ and pressure $P_t$ at the inner edge separating the liquid core and the solid crust of a neutron star. 
A theoretical framework that includes the thermodynamic method, relativistic nuclear energy density functionals and the quasiparticle random-phase 
approximation is employed in a self-consistent calculation of $(n_t,P_t)$ and collective excitations in nuclei. 
The covariance analysis shows that properties of charge-exchange dipole transitions, isovector giant dipole and quadrupole resonances and pygmy dipole transitions
are correlated with the core-crust transition density and pressure. A set of relativistic nuclear energy density functionals, characterized by systematic variation of the density dependence of the symmetry energy of nuclear matter, is used to constrain possible values for $(n_t,P_t)$. By comparing the calculated excitation energies of giant resonances, energy weighted pygmy dipole strength, and dipole polarizability with available data, we obtain the weighted average values: $n_t = 0.0955 \pm 0.0007$ fm$^{-3}$ and 
$P_t = 0.59 \pm 0.05$ MeV fm$^{-3}$.
\end{abstract}
\maketitle
\date{today}
%%%%%%%%%%%%%%%%%%%%%%%%%%%%%%%%%%%%%%%
%
%
The composition of the crust of a neutron star presents an interesting challenge both for 
nuclear structure physics as well as for nuclear astrophysics. A solid crust of 
$\approx 1$ km thickness and composed of  
nonuniform neutron-rich matter is located above a liquid core \cite{Han.07}. 
The crust presents an interface between observable surface phenomena and the
invisible core of the star, and its structure can be related to interesting effects, such as
glitches in the rotational period of pulsars, thermal relaxation after matter accretion, 
quasi periodic oscillations 
and anisotropic surface cooling~\cite{Lat.00}. The inner crust comprises the
region from the density at which neutrons drip out from nuclei,
to the inner edge separating the solid crust from the homogeneous
liquid core. At the inner edge, in fact,
a phase transition occurs from high-density homogeneous
matter to inhomogeneous matter at lower densities. While the density at 
which neutrons drip out from nuclei is rather
well determined, the transition density at the inner edge is
much less certain because of insufficient knowledge of the
equation of state of neutron-rich nuclear matter. 

A number of theoretical studies
have shown that the core-crust transition density and pressure are highly
sensitive to the poorly constrained density dependence of the nuclear matter 
symmetry energy~\cite{Hor.01,Mou.10,Xu.09,Duc.11}. The symmetry energy that governs
the composition of the neutron star crust, also determines the thickness of the neutron-skin 
$r_{np} = r_{n} - r_{p}$
in finite nuclei. In a model study of Ref.~\cite{Hor.01} an inverse correlation was found between 
the liquid-to-solid phase transition density for neutron-rich matter and the neutron-skin 
thickness of $^{208}$Pb.
Additional correlations between $r_{np}$ and neutron star properties have also 
been investigated~\cite{Ste.05}, including neutron star radii~\cite{Car.03}, the threshold density 
at the onset of the direct Urca process~\cite{Hor.02}, and the crustal moment of inertia~\cite{Ste.05,Fat.10}.
Correlations between $r_{np}$ and a variety of neutron star properties have recently been studied 
using covariance analysis based on relativistic energy density functional~\cite{Fat.12}.
As pointed out by Horowitz and Piekarewicz~\cite{Hor.01}, an accurate measurement
of the neutron radius in $^{208}$Pb by means of parity-violating electron scattering may have important 
implications for the structure of the crust of neutron stars. The parity radius experiment
(PREX) has recently provided the first model-independent evidence for the neutron skin 
in $^{208}$Pb~\cite{Abr.12}. However, since the experimental uncertainty of the 
PREX neutron skin thickness is very large $(r_{np}=0.33^{+0.16}_{-0.18}$ fm), it will be 
useful to explore additional experimental constraints for neutron star properties. Here we 
focus on collective nuclear excitations that correlate with $r_{np}$ and provide constraints
on the symmetry energy.

The purpose of this Letter is to analyze possible relationships between collective excitation
modes in finite nuclei and properties of the crust of a neutron star. Of particular importance
are the liquid-to-solid transition density and pressure, that is, quantities that determine the 
inner region of the crust. Recent experimental studies of giant resonances, pygmy dipole 
resonances and other modes of excitation in nuclei, yielded 
a wealth of data that constrain the nuclear symmetry energy and
neutron skin thickness~\cite{Sav.13}.  Since there is a direct 
relation between the liquid-to-solid transition density and the neutron radius of 
$^{208}$Pb~\cite{Hor.01}, one expects that an analysis of the collective response of finite 
nuclei could also have important implications 
for the structure of the crust of neutron stars. This assumption will be put to the test
in a theoretical framework that includes the thermodynamic method, relativistic nuclear energy density
functionals and covariance analysis.

To determine the liquid-to-solid transition density for neutron-rich matter, the usual 
approach is to find the density at which the uniform liquid becomes unstable
against small-amplitude density fluctuations, indicating the formation of nuclear clusters.
In this way a lower bound to the true transition density $n_t$ is obtained~\cite{Pet.95}. 
The procedures used to determine $n_t$ include  the dynamic method~\cite{Pet.95,Duc.07,Xu.09},  
the thermodynamic method~\cite{Lat.07,Kub.07,Wor.08,Mou.10}, and the
random-phase approximation (RPA) \cite{Hor.01}. For the purpose of the present study the
thermodynamic method will be employed. The constraint that determines 
the transition density is given by the inequality \cite{Lat.07,Mou.10} 
\begin{eqnarray}
C(n) = n^2\frac{{\rm d}^2V}{{\rm d}n^2}+2n\frac{{\rm
d}V}{dn} +  (1-2x)^2\left[ n^2\frac{{\rm d}^2 E_{sym}}{{\rm
d}n^2} \right. \nonumber \\ 
\left. +  2n\frac{{\rm d}E_{sym}}{{\rm d}n}-2\frac{1}{E_{sym}}\left(n
\frac{{\rm d}E_{sym}}{{\rm d}n} \right)^2 \right]>0, \label{K-I}
\end{eqnarray}
where $n$, $V$, $x$ and $E_{sym}$, denote  the baryon density, the energy per particle of 
symmetric nuclear matter, the proton fraction, and the symmetry energy, respectively. The
transition density $n_t$ is determined by solving the equation $C(n_t)=0$, 
and the corresponding transition pressure reads
$P_t(n_t,x_t)=P_b(n_t,x_t)+P_e(n_t,x_t)$, where $P_b$, $P_e$ are the baryon and electron
contributions, respectively. $x_t$ denotes the proton fraction that corresponds to $n_t$, and
is computed using the condition of $\beta$-equilibrium~\cite{Mou.10}.
For the analysis of correlations between the transition density and pressure $(n_t,P_t)$ and  
observables that characterize collective excitations in finite nuclei, we consistently 
employ a relativistic nuclear energy density functional (RNEDF) 
to compute the energy per particle of symmetric nuclear matter and the symmetry
energy, and in the RPA calculation of strength functions in finite nuclei.  In this work the 
universal RNEDF with density-dependent meson-nucleon couplings \cite{Nik.02} is used, 
and excitations in spherical nuclei are analyzed in the 
relativistic quasiparticle random phase approximation (RQRPA)~\cite{PVKC.07}. The density dependence of
the symmetry energy can be expressed in terms of coefficients of the Taylor expansion around 
nuclear matter saturation density $n_0$:
\begin{equation}
E_{sym}(n)=E_{sym}(n_0)+L\left(\frac{n-n_0}{3n_0}\right)+...
\end{equation}
where $E_{sym}(n_0) \equiv J$ is the symmetry energy at saturation, 
and $L$ denotes the slope parameter.
It has been shown that the parameters $J$,$L$ correlate not only with the neutron-skin thickness of  
nuclei~\cite{Fur.02,Cen.09}, but also with neutron star properties~\cite{Fat.11,Fat.12}.

\begin{figure}
\centering
\includegraphics[scale=0.35,angle=0]{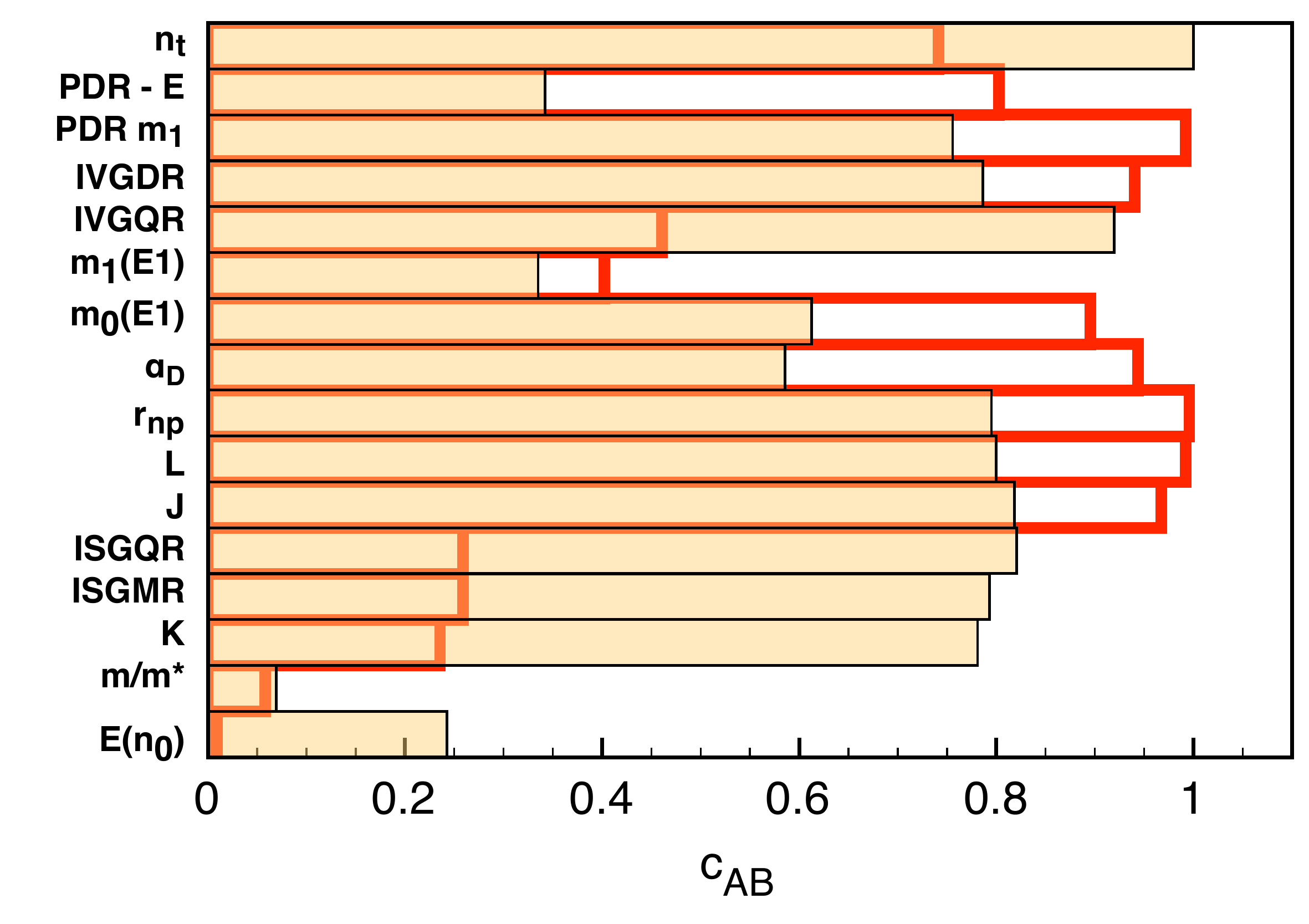}
\caption{(Color online). Pearson product-moment correlation coefficients between 
the density $n_t$ (shaded bars - thin lines) at the phase transition between the liquid 
core and the solid crust of a neutron star, and various observables of collective 
excitations for $^{208}$Pb and nuclear matter properties. The corresponding correlation coefficients 
for the pressure $P_t$ are displayed by thick (red) bars.}
\label{fig1}
\end{figure}

To assess the information content on the neutron star liquid-to-solid transition density and pressure,
that is carried by various observables that characterize collective excitations in nuclei, 
the RNEDF and covariance analysis~\cite{Rei.10} are
used to calculate correlations between quantities of interest. Covariance analysis is the least biased 
and most exhaustive approach to identify correlations between physical observables~\cite{Rei.10,Fat.12}.
For the purpose of a covariance analysis the DDME-min1 parameterization of the RNEDF 
has been developed by fitting to ground-state data, that is, binding energies, charge radii, 
diffraction radii, and surface thickness of 17 spherical nuclei, from $^{16}$O to $^{214}$Pb~\cite{Xav.14}.
Figure~\ref{fig1} shows the corresponding Pearson product-moment correlation 
coefficients~\cite{Rei.10} between the neutron star transition density $n_t$ (and pressure $P_t$ ) 
and various quantities that characterize nuclear matter and finite nuclei. The correlation coefficients
are obtained in a consistent implementation of the RNEDF and thermodynamic model, using the 
DDME-min1 parameterization. The following equilibrium nuclear matter properties are
included:  the binding energy at saturation density $E(n_0)$, the effective mass $m/m^*$, the incompressibility $K$, the symmetry energy $J$ and slope of the symmetry energy at saturation $L$ (Eq.(2)). Characteristic quantities of various modes of excitation of $^{208}$Pb are also taken into consideration: the excitation energies of the isoscalar giant monopole resonance (ISGMR) and isoscalar giant quadrupole resonance (ISGQR), the dipole polarizability 
$(\alpha_D)$, the overall isovector dipole transition strength $(m_0)$ and the respective energy-weighted dipole transition strength $(m_1)$, the excitation energies of the isovector giant quadrupole resonance (IVGQR) and isovector giant dipole resonance (IVGDR), the $m_1$ moment (PDR $m_1$) and excitation energy (PDR - E) of the pygmy dipole strength function. In addition, the neutron-skin thickness $(r_{np})$ in $^{208}$Pb is also included. 
The present covariance analysis confirms the strong linear correlation between $n_t$ and the neutron-skin thickness of $^{208}$Pb, as already shown in Ref.~\cite{Hor.01}, and also displays a similar correlation between 
$P_t$ and $r_{np}$. The results shown in Fig.~\ref{fig1} also indicate that collective excitations in finite nuclei 
are strongly correlated to the neutron star properties $n_t$ and $P_t$. To constrain possible values of ($n_t$,$P_t$), of particular interest are observables that simultaneously correlate with both quantities. These include the overall isovector dipole transition strength $m_0$, the IVGDR and IVGQR excitation energies, the PDR energy weighted transition strength, and the dipole polarizability.

\begin{figure}
\centering
\includegraphics[scale=0.35,angle=0]{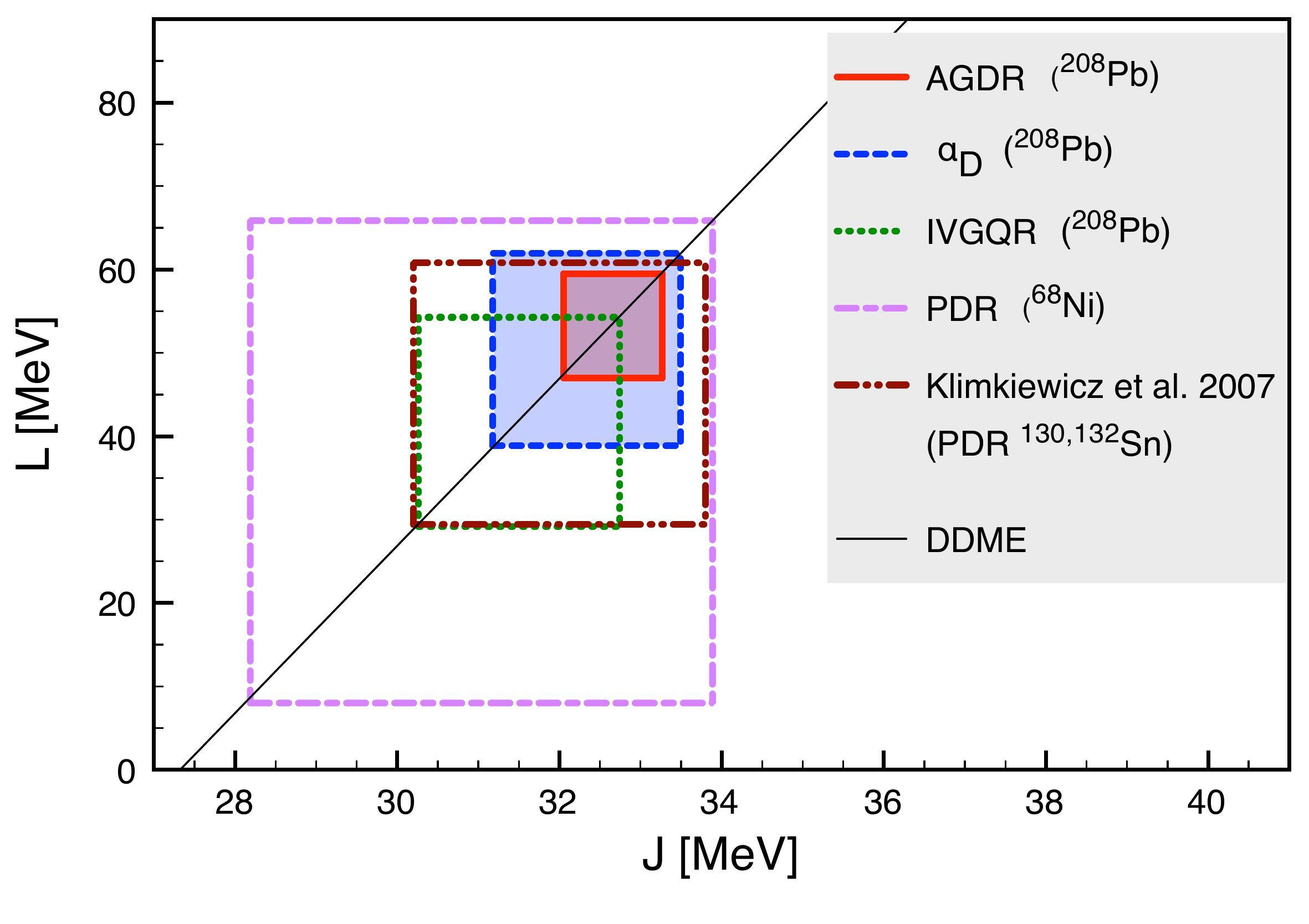}
\caption{(Color online). Constraints of the symmetry energy at saturation $J$ and the
slope parameter $L$, obtained from a comparison of RNEDF results 
and data on AGDR~\cite{Kra.13} and IVGQR~\cite{Hen.11} excitation energies ($^{208}$Pb), 
the dipole polarizability $\alpha_D$ of $^{208}$Pb~\cite{Tam.11}, and the PDR energy weighted 
strength ($^{68}$Ni~\cite{Wie.09}, $^{130,132}$Sn~\cite{Kli.07}).}
\label{fig2}
\end{figure}

Figure~\ref{fig1} shows that the liquid-to-solid transition density and pressure are also correlated with the 
symmetry energy coefficients $J$ and $L$. It is, therefore, interesting to analyze how various 
excitation modes in nuclei, that limit possible values of $r_{np}$, provide constraints on the 
density dependence of the symmetry energy. Similar studies have recently been
performed for different modes of excitation using the framework of energy density functionals 
(e.g. Refs.~\cite{Car.10,Tsa.12,Maz.13}).
In the present analysis a consistent set of RNEDFs that span a range of values $J=30-38$ MeV and
$L=30-110.8$ MeV~\cite{Vre.03}, is employed in a calculation of collective excitations.
The set of RNEDFs was adjusted to accurately reproduce nuclear-matter properties, binding energies 
and charge radii of a standard set of spherical
nuclei, but with constrained values for the symmetry energy $J$ and slope parameter $L$ \cite{Vre.03}. 
These functionals were recently used to constrain the density dependence of the 
nuclear symmetry energy and the neutron-skin thickness 
from the observed pygmy dipole strength ($^{130,132}$Sn)~\cite{Kli.07}, and the anti-analog giant
dipole resonance ($^{208}$Pb)~\cite{Kr.14}.
By performing self-consistent relativistic mean-field calculations for the nuclear ground states, and the 
corresponding RQRPA for collective excitations, we have computed 
the AGDR and IVGQR excitation energies in $^{208}$Pb, the dipole polarizability $\alpha_D$ of $^{208}$Pb, 
and the PDR transition strength in $^{68}$Ni. For the set of RNEDFs, linear correlations are established 
between the calculated characteristics of collective excitations and the symmetry energy $J$
and slope parameter $L$, in agreement with the results of the covariance analysis shown in Fig.~\ref{fig1}. 
These correlations, together with the corresponding experimental results on the excitation strengths and 
energies \cite{Kr.14,Hen.11,Tam.11,Wie.09}, provide independent constraints on $J$ and $L$, 
shown in Fig.~\ref{fig2}. For comparison we also include the results of a previous study that was 
based on the same set of RNEDFs, but used data on the PDR in $^{130,132}$Sn~\cite{Kli.07}. 
Figure~\ref{fig2} shows that all calculated excitation properties consistently constrain possible values of 
$J$ and $L$, with differences attributed to variations of the experimental uncertainties. 
It is interesting to note that all results overlap in a narrow region of the $(J,L)$ plane. The 
weighted average yields $J = 32.5 \pm 0.5$ MeV
and $L = 49.9 \pm 4.7$ MeV. More accurate experimental results would, of course, further reduce
the uncertainties shown in Fig.~\ref{fig2}. 

\begin{figure}
\centering
\includegraphics[scale=0.35,angle=0]{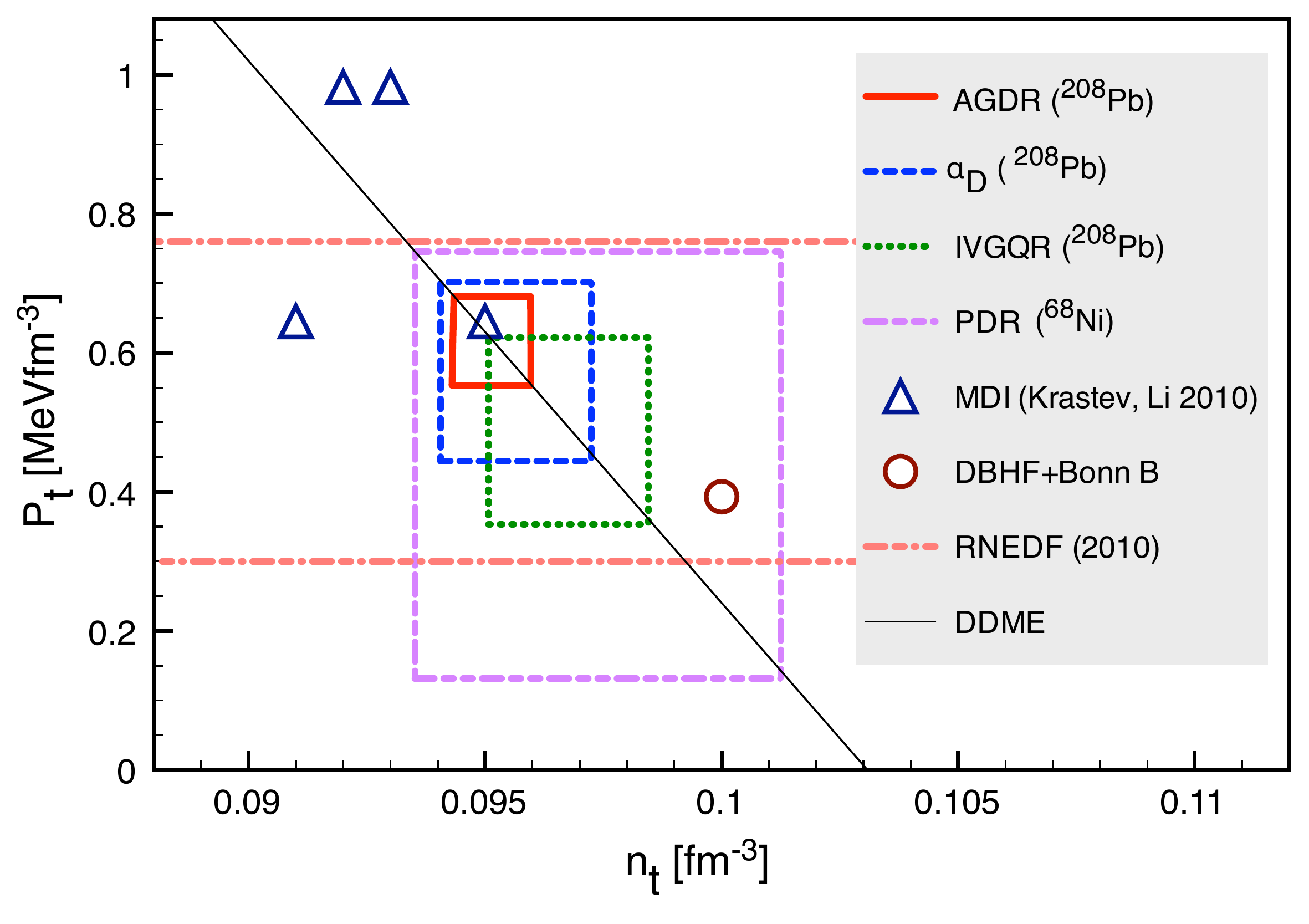}
\caption{(Color online).The liquid-to-solid transition pressure $P_t$ for neutron-rich matter as a
function of the transition density $n_t$. The rectangles correspond to the 
values of ($P_t$,$n_t$) calculated using the RNEDF and experimental data for 
AGDR~\cite{Kr.14} and IVGQR~\cite{Hen.11} excitation energies ($^{208}$Pb), the 
dipole polarizability $\alpha_D$ ($^{208}$Pb)~\cite{Tam.11}, and the PDR energy weighted
strength ($^{68}$Ni)~\cite{Wie.09}. Results obtained by several other methods: the MDI interactions \cite{Xu.09, Kra.13}, DBHF+Bonn B 
calculations \cite{Sam.09}, and the self-consistent RNEDF calculation \cite{Mou.10}, are
shown for comparison.}
\label{fig3}
\end{figure}

In the next step we use the same set of RNEDFs to compute the liquid-to-solid transition density and 
pressure in the thermodynamic approach of Eq.~(\ref{K-I}). In Fig.~\ref{fig3} the transition pressure 
$P_t$ is plotted as a function of the transition density $n_t$, and we notice the particular linear dependence 
predicted by the DDME set of relativistic functionals. The rectangles denote the values of $P_t$ and $n_t$, 
that is, the corresponding energy density functionals, that in a consistent RQRPA calculation 
reproduce data on collective excitations within experimental uncertanities: the AGDR~\cite{Kr.14} and 
IVGQR~\cite{Hen.11} excitation energies ($^{208}$Pb), the 
dipole polarizability $\alpha_D$ ($^{208}$Pb)~\cite{Tam.11}, and the PDR energy weighted
strength ($^{68}$Ni)~\cite{Wie.09}. One notices that collective excitations provide rather stringent 
constraints on the possible values of $P_t$ and $n_t$, and there is even a small region in the 
$(P_t,n_t)$ plane in which all constraints overlap. Obviously, more accurate measurements of 
charge-exchange modes and pygmy dipole strength are necessary to reduce current  
uncertainties but, nevertheless, the weighted average from the present analysis yields 
$n_t = 0.0955 \pm 0.0007$ fm$^{-3}$ and 
$P_t = 0.59 \pm 0.05$ MeV fm$^{-3}$. 

For comparison, Fig.~\ref{fig3} also includes constraints 
on $(P_t,n_t)$ obtained by other methods, based on modified Gogny (MDI) interactions~\cite{Xu.09, Kra.13},
Dirac-Brueckner-Hartree-Fock (DBHF) calculations~\cite{Sam.09}, and RNEDF calculations supplemented
with constraints from the empirical range for the slope parameter $L$ and neutron-skin thickness in  
Sn isotopes and $^{208}$Pb~\cite{Mou.10}. We note that a previous study based on the 
$A18+\delta v+UIX^*$ interaction predicted a somewhat lower value for the transition density:  
$n_t$=0.087 fm$^{-3}$~\cite{Akm.98}, whereas the constraints  
obtained in the present analysis are consistent with the result based on the 
nonrelativistic microscopic equation of state of
Friedman and Pandharipande~\cite{Fri.81}: $n_t$=0.096 fm$^{-3}$~\cite{Lor.93}.

In conclusion, using a theoretical framework based on nuclear density functional theory, 
it is shown that the neutron star liquid-to-solid phase transition density
and pressure are correlated with properties of collective excitations in finite nuclei. 
A set of RNEDFs characterized by systematic variation of the density 
dependence of the symmetry energy, is used in a 
self-consistent RPA calculation of charge-exchange transitions, 
isovector giant quadrupole resonances, pygmy dipole 
transitions and dipole polarizability. A comparison with available data yields 
the following weighted average values for the symmetry energy at saturation 
$J = 32.5 \pm 0.5$ MeV, and the slope parameter $L = 49.9 \pm 4.7$ MeV. 
Using the thermodynamic method, the same set of RNEDFs 
predicts a particular linear dependence of the phase transition density $n_t$ and 
pressure $P_t$ at the inner edge between the liquid core
and the solid crust of a neutron star. By comparing the corresponding RPA 
theoretical values for the AGDR and 
IVGQR excitation energies ($^{208}$Pb), the 
dipole polarizability $\alpha_D$ ($^{208}$Pb), and the PDR energy weighted
strength ($^{68}$Ni) with available data, one obtains 
rather stringent constraints on the possible values: 
$n_t = 0.0955 \pm 0.0007$ fm$^{-3}$ and 
$P_t = 0.59 \pm 0.05$ MeV fm$^{-3}$. Of course, the present analysis 
has been based on a single family of RNEDFs with the aim to demonstrate 
the feasibility of the proposed theoretical approach. Further extensive
studies with a wider range of non-relativistic and relativistic functionals will provide more 
robust quantitative estimates of the liquid-to-solid transition density and pressure. This 
method crucially depends on experimental uncertainties of observables that characterize  
collective modes of excitation and, therefore, accurate measurements are necessary 
to further constrain the structure of the neutron star crust.

%
%------------------------------------------------------------------


\begin{thebibliography}{999}
\bibitem{Han.07} P. Haensel, A.Y. Potekhin, and D.G. Yakovlev, Neutron Stars 1: Equation of State
and Structure (Springer-Verlag, New York, 2007).
\bibitem{Lat.00} J.M. Lattimer and M. Prakash, Phys. Rep.  {\bf 333}, 121 (2000).
\bibitem{Hor.01} C. J. Horowitz, J. Piekarewicz, Phys. Rev. Lett.  {\bf 86}, 5647 (2001). 
\bibitem{Mou.10} Ch. C. Moustakidis, T. Nik\v si\'c, G. A. Lalazissis, D. Vretenar, and P. Ring, 
Phys. Rev. C  {\bf 81}, 065803 (2010).
\bibitem{Xu.09} J. Xu, L. W. Chen, B. A. Li, and H. R. Ma, Astrophys. J.  {\bf 697},
1549 (2009).
\bibitem{Duc.11} C. Ducoin, J. Margueron, C. Providencia, I. Vidana, Phys. Rev. C {\bf 83}, 045810 (2011).
\bibitem{Ste.05} A. W. Steiner, M. Prakash, J. M. Lattimer, and P. J. Ellis, Phys. Rep.  {\bf 411}, 325 (2005).
\bibitem{Car.03} J. Carriere, C. J. Horowitz, and J. Piekarewicz, Astrophys. J.  {\bf 593}, 463 (2003).
\bibitem{Hor.02} C. J. Horowitz and J. Piekarewicz, Phys. Rev. C  {\bf 66}, 055803 (2002).
\bibitem{Fat.10} F. J. Fattoyev and J. Piekarewicz, Phys. Rev. C  {\bf 82}, 025810 (2010).

\bibitem{Fat.12} F. J. Fattoyev and J. Piekarewicz, Phys. Rev. C  {\bf 86}, 015802 (2012).

\bibitem{Abr.12} S. Abrahamyan et al., Phys. Rev. Lett.  {\bf 108}, 112502 (2012).
\bibitem{Sav.13} D. Savran, T. Aumann, A. Zilges, Prog. Part. Nucl. Phys.  {\bf 70}, 210 (2013).

\bibitem{Pet.95} C. J. Pethick, D. G. Ravenhall, and C. P. Lorenz, Nucl. Phys. A  {\bf 584}, 675 (1995); F. Douchin and P. Haensel, Phys. Lett. B  {\bf 485}, 107 (2000).
\bibitem{Duc.07} C. Ducoin, Ph. Chomaz, and F. Gulminelli, Nucl. Phys. A  {\bf 789}, 403 (2007).

\bibitem{Lat.07} J. M. Lattimer and M. Prakash, Phys. Rep.  {\bf 442}, 109 (2007).
\bibitem{Kub.07}S. Kubis, Phys. Rev. C {\bf 76}, 025801 (2007).
\bibitem{Wor.08}A. Worley, P.G. Krastev, and B.A. Li, Astrophys. J. {\bf 685}, 390 (2008).
\bibitem{Nik.02} T. Nik{\v{s}}i{\'{c}}, D. Vretenar, P. Finelli, and P. Ring,
	 	Phys. Rev. C  {\bf 66}, 024306 (2002). 
\bibitem{PVKC.07} N. Paar, D. Vretenar, E. Khan, and G. Col\`o,
Rep. Prog. Phys.  {\bf 70}, 691 (2007).
\bibitem{Fur.02} R. J. Furnstahl, Nucl. Phys. A  {\bf 706}, 85 (2002).
\bibitem{Cen.09} M. Centelles, X. Roca-Maza, X. Vinas, and M. Warda, Phys. Rev. Lett.  {\bf 102}, 122502 (2009).
\bibitem{Fat.11} F. J. Fattoyev and J. Piekarewicz, Phys. Rev. C  {\bf 84}, 064302 (2011).


\bibitem{Rei.10} P.-G. Reinhard and W. Nazarewicz, Phys. Rev. C  {\bf 81}, 051303(R) (2010).
\bibitem{Xav.14} X. Roca-Maza, N. Paar, G. Col\'o, to appear in J. Phys. G (2014).
\bibitem{Car.10} A. Carbone, G. Col\' o, A. Bracco et al., Phys. Rev. C  {\bf 81}, 041301(R)(2010).
\bibitem{Tsa.12} M. B. Tsang et al., Phys. Rev. C  {\bf 86}, 015803 (2012).
\bibitem{Maz.13} X. Roca-Maza, M. Brenna, B. K. Agrawal et al., Phys. Rev. C  {\bf 87}, 034301 (2013).
\bibitem{Vre.03} D. Vretenar, T. Nik\v si\'c, P. Ring, Phys. Rev. C  {\bf 68}, 024310 (2003).

\bibitem{Kli.07} A. Klimkiewicz et al., Phys. Rev. C  {\bf 76}, 051603(R) (2007).
\bibitem{Kr.14} A. Krasznahorkay et al.,  arXiv:1311.1456 (2014).
%
\bibitem{Hen.11} S. S. Henshaw et al., Phys. Rev. Lett.  {\bf 107}, 222501 (2011).
\bibitem{Tam.11} A. Tamii et al., Phys. Rev. Lett.  {\bf 107}, 062502 (2011).
\bibitem{Wie.09} O. Wieland, A. Bracco, F. Camera et al., Phys. Rev. Lett.  {\bf 102}, 092502 (2009).
\bibitem{Kra.13} P. G. Krastev and B. A. Li, arXiv:1001.0353 [astro-ph.SR].
\bibitem{Sam.09} F. Sammarruca and P. Liu, Phys. Rev. C  {\bf 79}, 057301 (2009);
F. Sammarruca, arXiv:1002.0146 [nucl-th].
\bibitem{Akm.98} A. Akmal, V. R. Pandharipande, and D. G. Ravenhall, Phys. Rev. C  {\bf 58}, 1804 (1998).
\bibitem{Fri.81}  B. Friedman and V. R. Pandharipande, Nucl. Phys. A  {\bf 631}, 502 (1981).
\bibitem{Lor.93} C. P. Lorenz, D. G. Ravenhall, and C. J. Pethick, Phys. Rev. Lett.  {\bf 70}, 379 (1993).
%
\end{thebibliography}
\end{document}